\PassOptionsToPackage{english}{babel}
\documentclass[useAMS,usenatbib]{mn2e}

\usepackage[english]{babel}
\usepackage{color}
\usepackage{graphicx}
\usepackage{breqn}
\usepackage{amssymb}
\usepackage{tabu}
\usepackage[draft]{hyperref}
\usepackage{caption}
\usepackage{booktabs,caption}
\usepackage{subcaption}
\title[Observability of CPDs]{Observability of Forming Planets and their Circumplanetary Disks II. -- SEDs and Near-Infrared Fluxes}
\author[Szul\'agyi et al.]{
\parbox{\textwidth}{
J. Szul\'agyi$^{1}$\thanks{E-mail:
judit.szulagyi@uzh.ch}, C. P. Dullemond$^{2}$, A. Pohl$^{3,2}$ \&  S. P. Quanz$^{4}$\\
}\vspace{3mm}\\
$^{1}$ Center for Theoretical Astrophysics and Cosmology, Institute for Computational Science, University of Z\"urich,\\ Winterthurerstrasse 190, CH-8057 Z\"urich, Switzerland\\
$^{2}$ Heidelberg University, Center for Astronomy, Institute of Theoretical Astrophysics, Albert-Ueberle-Str. 2, 69120 Heidelberg, Germany\\
$^{3}$ Max-Planck-Institute for Astronomy, K\"onigstuhl 17, D-69117 Heidelberg, Germany\\
$^{4}$ ETH Z\"urich, Institute for Particle Physics and Astrophysics, Wolfgang-Pauli-Strasse 27, CH-8093, Z\"urich, Switzerland\\
}

\begin{document}

\date{Accepted XX. Received XX; in original form 2018 Aug 2}

\pagerange{\pageref{firstpage}--\pageref{lastpage}} \pubyear{2018}

\maketitle

\label{firstpage}

\begin{abstract}
Detection of forming planets means detection of the circumplanetary disk (CPD) in reality, since the planet is still surrounded by a disk at this evolutionary stage. Yet, no comprehensive CPD modeling was done in near-infrared wavelengths, where high contrast imaging is a powerful tool to detect these objects. We combined 3D radiative hydrodynamic simulations of various embedded planets with RADMC-3D radiative transfer post-processing that includes scattering of photons on dust particles. We made synthetic images for VLT NaCo/ERIS in the Ks, L', M' bands as well as examined the spectral energy distributions (SEDs) of disks between 1 $\mu m$ and 10 cm. 
We found that the observed magnitudes from the planet's vicinity will mostly depend on the CPD parameters, not on the planet's. The CPD is 20-100x brighter than the embedded planet in near-IR. We also show  how the CPD parameters, e.g. the dust-to-gas ratio will affect the resulting CPD magnitudes. According to the SEDs, the best contrast ratio between the CPD and circumstellar disks is in sub-mm/radio wavelengths and between 8-33 microns in case if the planet opened a resolvable, deep gap ($\ge 5 \rm{M_{Jup}}$), while the contrast is particularly poor in the near-IR. Hence, to detect the forming planet and its CPD, the best chance today is targeting the sub-mm/radio wavelengths and the 10-micron silicate feature vicinity. In order to estimate the forming planet's mass from the observed brightness, it is necessary to run system specific disk modeling.

\end{abstract}


\section{Introduction}

The observations of forming planets are the keys to understand the giant planet formation processes. For this purpose, high-contrast imaging is often used, mainly in the near-infrared (near-IR). This way hot, point-like sources can be detected within the natal circumstellar disk (CSD), for which a natural, possible explanation is a protoplanet. There are a handful of such detections, the most obvious example being the recent discovery of PDS 70b \citep{Keppler18}. Furthermore, there is one candidate around MWC 758 \citep{Reggiani17}, three candidates around LkCa15 \citep{KI12,Sallum15}, two potential forming planets around HD100546 \citep{Britain14,Quanz15,Currie14}, two around HD169142 \citep{Reggiani14,Osorio14}, even though the origins of these hot spots are in some cases debated, as they could also be circumstellar disk features \citep{Follette17,Rameau17}. 

While there is still ongoing gas accretion to the forming giant planet, it is surrounded by its own disk, the so-called circumplanetary disk (CPD; e.g. \citealt{QT98,Lubow99,Kley99,LD06,AB09,AB12}. Roughly speaking, the lifetime of the gaseous CPD is nearly equal to the gaseous circumstellar disk (CSD) lifetime, because the CPD is constantly fed from the natal protoplanetary disk \citep{Szulagyi14,Szulagyi17gap}. As the gaseous CSD is dissipating, the transport of matter onto the CPD also ceases, the two disks surface density evolves together \citep{Szulagyi17gap}. When there is no more gas left in the CSD near the location (the feeding zone) of the planet and CPD, the feeding stops. As the CPD mass is very small even during the class II phase of the CSD ($\sim 10^{-3} M_{planet}$ \citealt{Szulagyi17gap}), at the time when the CSD disappears, the CPD is even lighter. This small mass ($< 10^{-3} M_{planet}$) will quickly disappear, partially by being accreted onto the gas giant, and by being dissipated by other processes (e.g. viscous spreading). Therefore, detecting a forming planet is actually detecting its gaseous CPD.

There is no comprehensive study made to date about detecting CPDs in the near-IR. \citet{QT98} performed analytical calculations which revealed that the CPD should be bright in the near-mid IR. \citet{Wolf18} did a parameter study using solely a radiative transfer algorithm to understand whether hot-spots of protoplanets can be detected in the mid-IR with the upcoming VLTI/MATISSE instrument. This study however did not include CPD, given there were no hydrodynamic simulations made. \citet{Eisner15} have studied the SEDs of potential CPDs by downscaling the stellar spectrum to a colder black-body. \citet{Zhu15} computed the disk spectrum via the emission from the atmosphere of a viscous, geometrically thin, optically thick accretion disk with a constant mass accretion rate. Due to the large debate in the community about whether the detected hot spots are real planets or disk features, here we address the observability of CPDs in the near-IR. We also provide their spectral energy distribution (SED) between 1 micron and 10 cm. We did a thorough study including temperature-included (i.e. radiative) 3D gas hydrodynamic simulations, that were post-processed with a wavelength-dependent radiative transfer algorithm that includes scattering on dust particles. Finally, we convolved the synthetic images with a Gaussian kernel to mimic observational data, using a point-spread function (PSF) for VLT/NaCo and VLT/ERIS. Because the brightness strongly depends on the opacity, we carefully made an opacity table that was consistently used both in the hydrodynamic simulations and in the post-processing radiative transfer runs. Our simulations include a circumstellar disk with an embedded planet at 50 AU. Between the different runs, we changed the mass of the planet between Saturn, Jupiter, 5-Jupiter and 10-Jupiter mass. These runs were made with the same surface density and same viscosity, because a full parameter space cannot be explored with these computationally expensive simulations. 

In the first paper of this series, we looked at the CPD observability at sub-mm/radio wavelength \citep{Szulagyi17alma}, and in next one we plan to address the scattered light imaging with polarization for instruments like VLT/SPHERE.

\section[]{Methods}
\label{sec:numerical}

As a first step we carried out radiative hydrodynamic simulations of the protoplanetary disk with a forming planet embedded within (Sect. \ref{sec:hydro}). Then we used the RADMC-3D radiative transfer tool to create wavelength-dependent images of the systems, including photon scattering on the dust particles (Sect. \ref{sec:radmc3d}). Finally, we convolved the images with a diffraction limited PSF for the VLT/NaCo and VLT/ERIS instruments \citep{Davies18} (Sect. \ref{sec:convol}) and calculated the apparent magnitudes of the circumplanetary disks and of the embedded planets from these images. We also created SEDs between 1 micron and 10 cm to understand which wavelength-range is the best to detect the CPDs. 

\subsection{Hydrodynamic Simulations}
\label{sec:hydro}

We performed three-dimensional radiative hydrodynamic simulations of circumstellar disks that have forming planets embedded in them with various planetary masses. The orbital distance of the planets from their star were 50 AU. We used the JUPITER code, that was developed by F. Masset and J. Szul\'agyi \citep{Szulagyi14,Szulagyi16a,Borro06} that not only solves Euler equations but also radiative transfer in the flux-limited diffusion approximation with the two-temperature approach \citep[e.g.][]{Commercon11,Bitsch14} that calculates the temperature field for the gas. The gas can heat up due to adiabatic compression, viscous heating and stellar irradiation, while it can cool through adiabatic expansion and radiative diffusion. The vicinity of the gas giant is heated up mainly by the accretion process \citep{Szulagyi16a}, as the gas tries to fall onto the planet, leading to adiabatic compression in this region. In a fast rotating (i.e. cold) circumplanetary disk, the viscous heating is also playing a role. The stellar irradiation, however, has no effect in the circumplanetary area. It only heats the atmosphere of the circumstellar disk, as the CPD is shielded from the stellar photons by the inner CSD. For observational predictions in the optical and near-IR, it is very important to simulate the circumstellar disk atmosphere (i.e. optically thin) regions, because here the photons can scatter multiple times before they get absorbed.

The Rosseland- and Planck-mean opacities used in the radiation-hydrodynamics code are constructed self-consistently from frequency dependent dust opacities computed with a version of the Mie code from \citet{BH84}. The dust consists of 40\% silicates, 40\% water and 20\% carbonaceous material \citep{Draine03,Zubko96,WB08}, assuming micron sized, spherical \& compact grains. The dust-to-gas ratio was set to 1\%, constant everywhere inside the simulation.  The opacity table accounts for the evaporation of the various dust species. We set the evaporation temperature for water, silicate and carbon to 170\,K, 1500\,K and 2000\,K respectively. Therefore, above 2000\,K the gas opacities play a role, that were taken from \citet{BL94}. We connected the different opacity regimes with splines to ensure smooth transitions. The hydrodynamic code uses Rosseland-mean opacities, more precisely the mass-weighted averages of the three dust species. In each cell of the hydrodynamic simulation the code uses the density and temperature of the given cell to look up what value of opacity it should use there. Then the flux-limited diffusion approximation algorithm finds the new temperature of the cell via an iterative method. In conclusion, even though the dust is not treated explicitly in the gas hydrodynamic simulation as a secondary fluid, its effect on the temperature of the disk is taken into account through the dust opacities (with the limit of assuming a constant dust-to-gas ratio of 1\%).

The star was assumed to have solar properties in the simulations and in the opacity table. Given that we were particularly interested in the planet's vicinity -- where high-resolution is necessary to get the temperatures correctly, we used mesh refinement in this region. This means that while the ring of the circumstellar disk has been simulated with a lower resolution (680 cells azimuthally over $2\pi$, 215 cells radially between 20 and 120 AU and 20 cells in the colatitude direction over 7.4 degrees opening angle from the midplane), the Hill-sphere of the planet is well resolved with four levels of refinement for these planetary masses and orbital separations. Each refinement doubles the resolution in each spatial direction, so the final resolution in the planet vicinity was approximately 0.029 AU. Even though in a usual circumstellar disk there is still some gas left within 20 AU from the star, this region has no effect on the circumplanetary disk, that is located at 50 AU from the star, hence we did not simulate this inner CSD region to save computational time.

The circumstellar disk ring had a mass of $\sim 10^{-2} \mathrm{M_{Sun}}$ with a surface density slope of -0.5 initially, which evolved due to the heating-cooling effects and the inclusion of the planets (i.e. gap-opening). We performed four different simulations with different planetary masses: a Saturn-mass gas giant, a Jupiter, a $5\mathrm{M_{Jup}}$, and a $10\mathrm{M_{Jup}}$ gas giant. The final planet masses were reached over 50 orbits through a smooth, sinusoidal function, to not perturb the gas flow very abruptly. Once this initial growth phase was over, the planet masses were kept fixed and the results were evaluated when steady state had been reached a few hundred orbits later. The planet point-mass was placed in the center of 8 cells, where we smoothed the gravitational potential as a usual and necessary technique in planet hydrodynamic simulations. The temperature values within the smoothing length (i.e. what we consider a planet) ranged between 1000 and 4000 K for the different planets, as the radiative module calculated. These values are consistent with the effective temperatures of Jupiter-like planets at the age of 1 Myr of planet interior \& evolution models \citep{Mordasini18}.

The equation of state in the hydrodynamic simulations was taken to be that of an ideal gas: $P=(\gamma-1)e$, where $\gamma=1.43$ adiabatic index connects the pressure $P$ with the internal energy $e$. The mean molecular weight was 2.3, corresponding to the solar abundances. The viscosity was a constant kinematic viscosity of $10^{-5} \mathrm{a_{p}}^2\Omega_p$, where $ \mathrm{a_{p}}$ is the semi-major axis and $\Omega_p$ denotes the orbital frequency of the planet.

\subsection{RADMC-3D post-processing}
\label{sec:radmc3d}

To create wavelength-dependent intensity images from the hydrodynamic simulations, we used the RADMC-3D \citep{Dullemond12}\footnote{\url{http://www.ita.uni-heidelberg.de/~dullemond/software/radmc-3d/}} radiative transfer tool. Even though the hydro simulations had a basic radiative transfer technique implemented, the results were not wavelength-dependent, because of the use of a Rosseland-mean-opacity. 

In the near-infrared, the scattering of photons on the dust particles is an important mechanism that significantly changes the observability of the circumplanetary disk and the planet. While this is of particular importance in the optically thin regions (i.e. in the CSD atmosphere), but a large atmosphere cannot be included into the hydrodynamic simulations. It is a known caveat that hydro solvers, especially the Riemann-solver used in our case, fail if the density and energy is too low in a couple of neighboring cells. The code slows down to a level that the computation basically stalls. Our experience is that below 7.5 degrees disk opening angle (that corresponds to roughly 4-5 pressure scale-heights) the computations are still safe. To be realistic for the photon scattering, therefore the disk atmosphere had to be extended by extrapolation for the RADMC-3D calculations. In the vertical direction, we fitted Gaussians to the density field, and extrapolated the low-density region so that instead of 40 cells in the co-latitude direction we had finally 100. In this region, we kept the temperature as in the last (optically thin) co-latitude cells to be sure that the stellar irradiation is still taken account. This means that the temperature in the disk atmosphere was constant with co-latitude and higher than the temperature in the bulk of the disk. For the scattering runs, we used $10^7$ photons and verified that this value was enough to reach convergence.

To be consistent with the hydrodynamic simulations and the opacity table we took the RADMC-3D parameters for the star, the dust and the disk identical to those of the hydrodynamical run. The dust-density files were created from the gas density (which is a good assumption as long as the dust grains are micron size and thus are strongly coupled to the gas), by multiplying the gas density in each cell with the dust-to-gas ratio. We assumed thermal equilibrium, hence we used the dust temperature to be equal to the gas temperature, while taking care of the evaporation of the water, silicates, and carbon. Therefore, in the RADMC-3D runs we used three dust species separated according to their evaporation temperatures:
\begin{enumerate}
\item {\bf Below 170 K:} the dust contains of a mixture of 40\% silicate, 40\% water and 20\% carbon. The dust density is 1\% of the gas density.
\item {\bf Between 170 K and 1500 K:} water has evaporated, hence the dust contains only silicates and carbon. In these cells the dust density is 0.6\% of the gas density.
\item {\bf Between 1500 K and 2000 K:} silicates have evaporated, hence the dust contains only carbon. The dust density is 0.2\% of the gas density.
\end{enumerate}
Above 2000 K even carbon evaporates, hence in these cells (e.g. very close to the planet) the dust-density was set to zero. While in the hydrodynamic simulations the gas opacity was adapted from \citet{BL94}, for the RADMC-3D run they cannot be used, as they are mean-opacities. Because it would be extremely difficult to obtain the frequency-dependent gas opacities \citep{Malygin14}, we simply assume that the gas is optically thin in the dust-free zones, even though this is not accurate everywhere.

The RADMC-3D image resolution was set to 1000 $\times$ 1000 pixels in each case to avoid resolution problems. The distance of the circumstellar disk was assumed to be 100 parsec for the calculations of the apparent magnitudes, and for the SEDs.

The SEDs were created with the same initial parameters and data files as the images (including the dust-temperature that is calculated by the hydrodynamic code). We ran the {\tt{sed}} command first for the  circumstellar disk with using the {\tt{fluxcons}} command to ensure flux conservation. To determine the CPD SEDs, we ran again the command but with the help of {\tt{zoomau}} to zoom into the CPD vicinity with a box-size of 5 AU in each direction from the planet. Due to the high computational cost of the SED runs (making 500 individual image within 1 micron and 10 cm wavelength-range), the amount of photons for scattering were set to "only" $10^4$, which already cost a few weeks of core-time. 

\subsection{Flux Determination}
\label{sec:convol}

\begin{figure*}
   \centering
\begin{tabular}{ccc}
  \hline
 Ks band (2.12 microns) & L' band (3.78 microns) & M band (4.76 microns) \\
  \hline
\\
\includegraphics[scale=0.35]{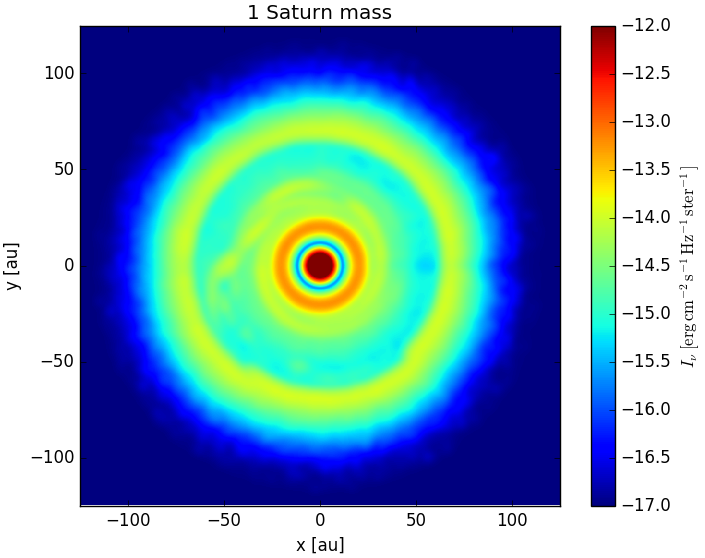} & \includegraphics[scale=0.35]{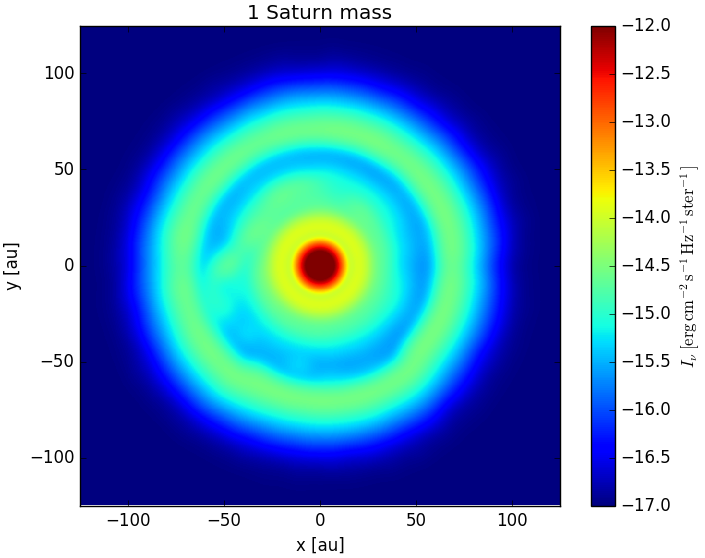} &  \includegraphics[scale=0.35]{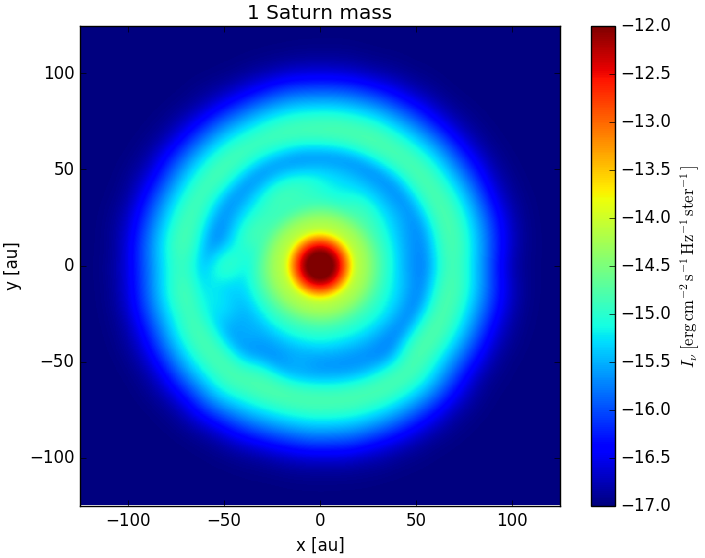}\\
\includegraphics[scale=0.35]{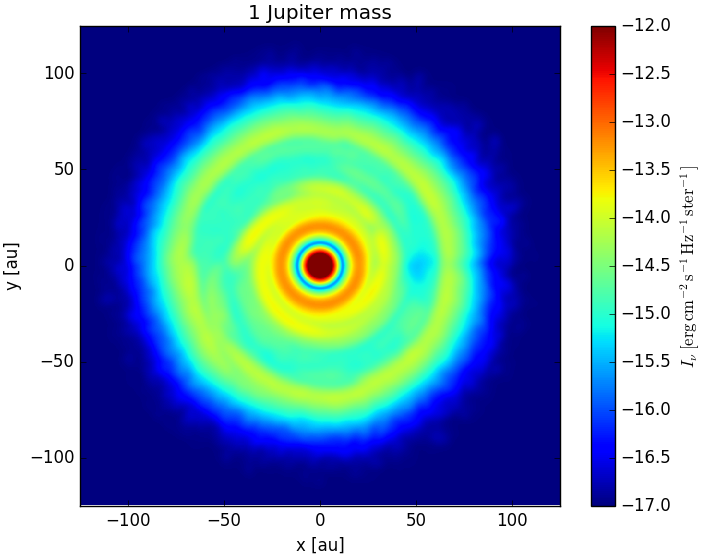} & \includegraphics[scale=0.35]{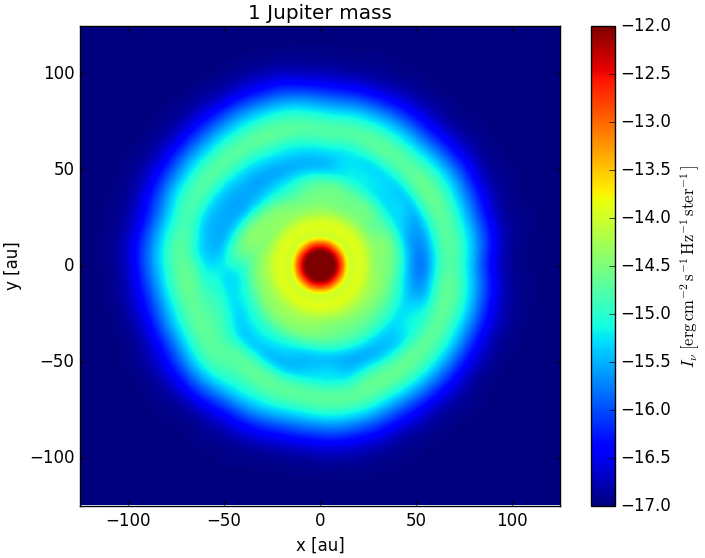} &  \includegraphics[scale=0.35]{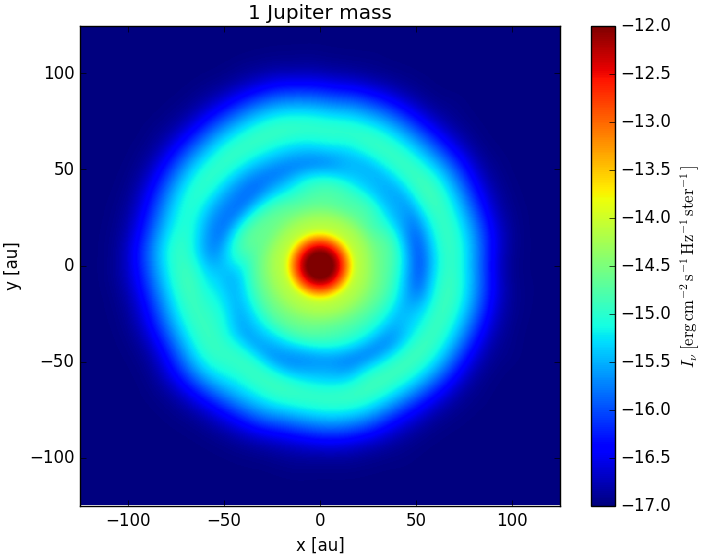}\\
\includegraphics[scale=0.35]{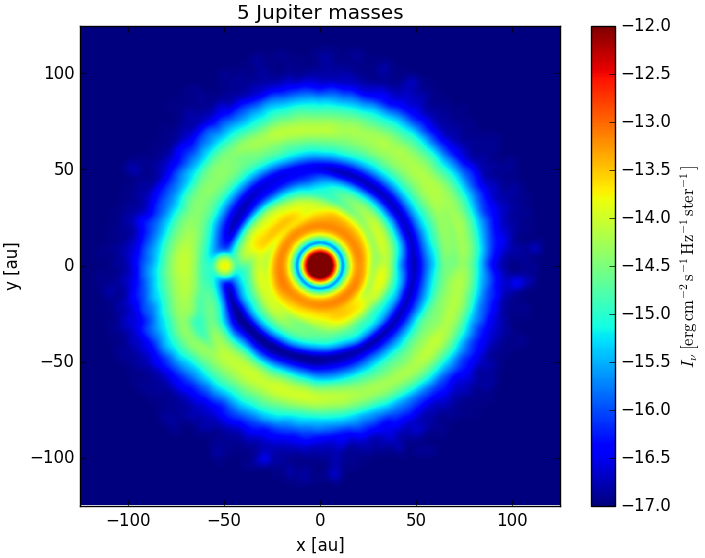} & \includegraphics[scale=0.35]{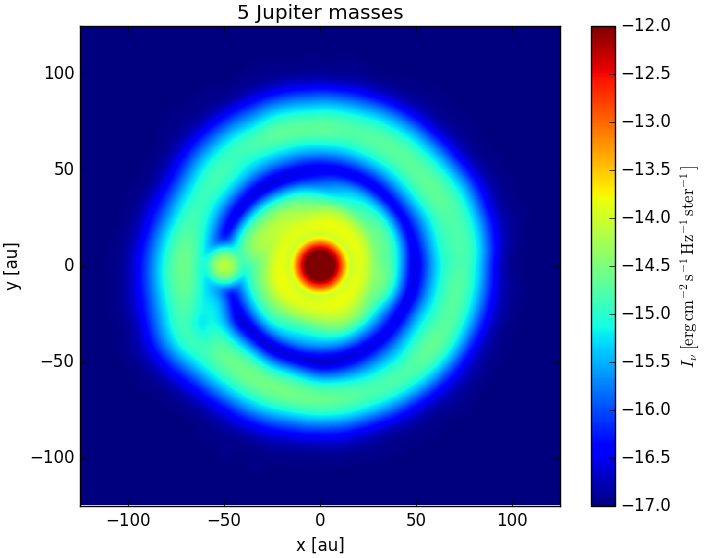} &  \includegraphics[scale=0.35]{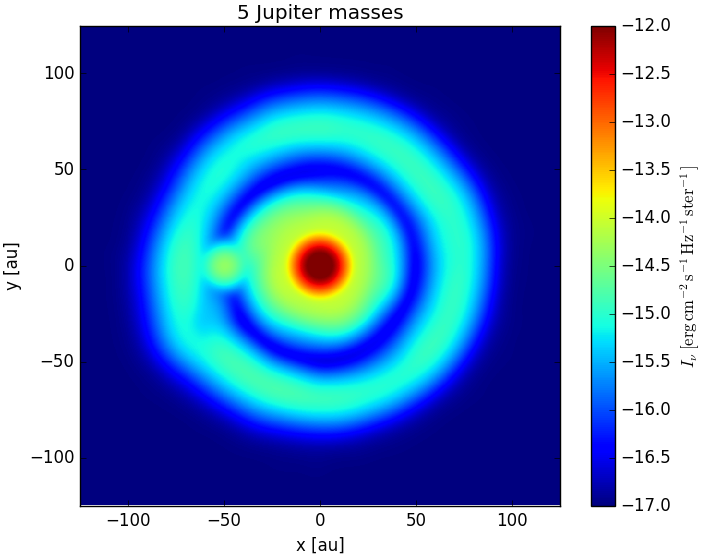}\\
\includegraphics[scale=0.35]{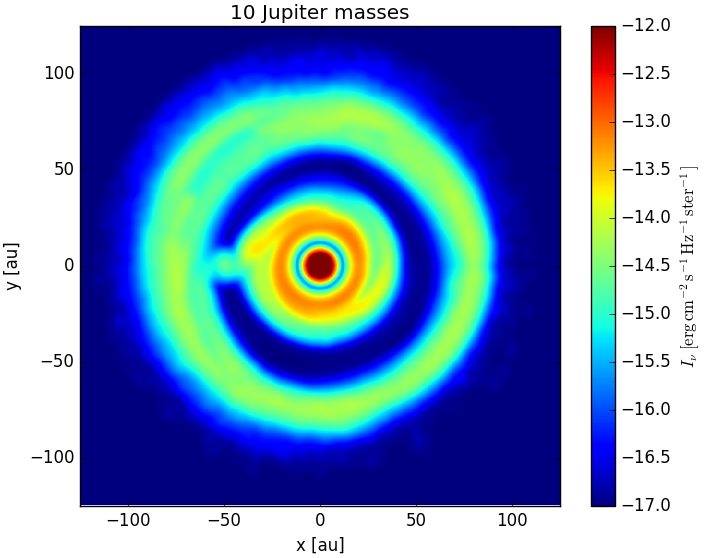} & \includegraphics[scale=0.35]{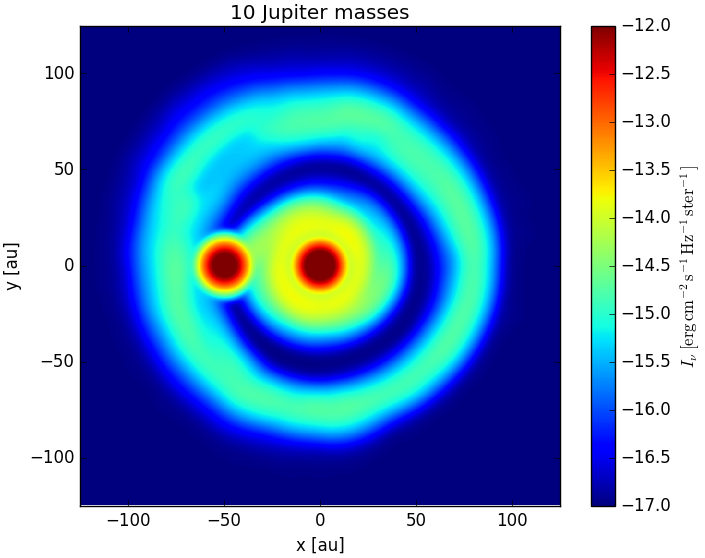} &  \includegraphics[scale=0.35]{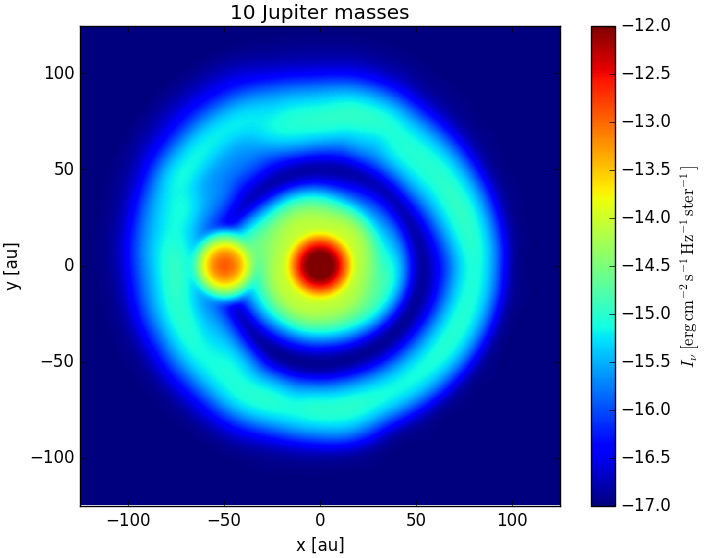}\\
\end{tabular}
    \caption{PSF-Convolved Synthetic Images (telescope diameter = 8m, distance = 100pc)}
    \label{fig:mocks} 
\end{figure*}

Given that we would like to provide realistic synthetic images, we choose the parameters of aperture and PSF according to an observation on HD100546 disk \citep{Quanz15} taken with the NaCo instrument at the Very Large Telescope. According to the diffraction limits in these wavelengths, we used for the FWHM of the Gaussian PSF 56 mas in Ks band (2.1 microns), 98 mas in L' band (3.8 microns) and 123 mas in M' band (4.8 microns). The aperture-radii were the same as in \citet{Quanz15}. The convolved RADMC-3D synthetic images are shown on Figure \ref{fig:mocks}. These images do not contain noise, so the presence of the atmosphere is not taken into account, because the noise will depend on the seeing during an actual observation. Because our models have a planet at a known location, it was obvious to determine the flux within one aperture centered on the planet. Given that there is significant contribution of the circumstellar disk to the final flux, we also determined the flux at the anti-planet location, i.e. the other side of the CSD with 180 degrees shift azimuthally. We then subtracted the flux of this CSD patch from the CPD flux. We also derived the embedded planet flux, as the peak flux within the aperture of the CPD. We converted the fluxes into apparent magnitudes at 100 pc distance. For this, the zero magnitude flux were taken as $\rm{F_{\nu}=653}$ Jy for Ks, 253 Jy for L', and 150 Jy for M' \footnote{\url{https://www.gemini.edu/sciops/instruments/midir-resources/imaging-calibrations/fluxmagnitude-conversion}}.

In the case of CPD we determined the SED at the planet vicinity using the {\tt{radmc3d sed zoomau}} commands.

\section{Results}

\subsection{SEDs}
\label{sec:seds}

\begin{figure*}
\includegraphics[width=18cm]{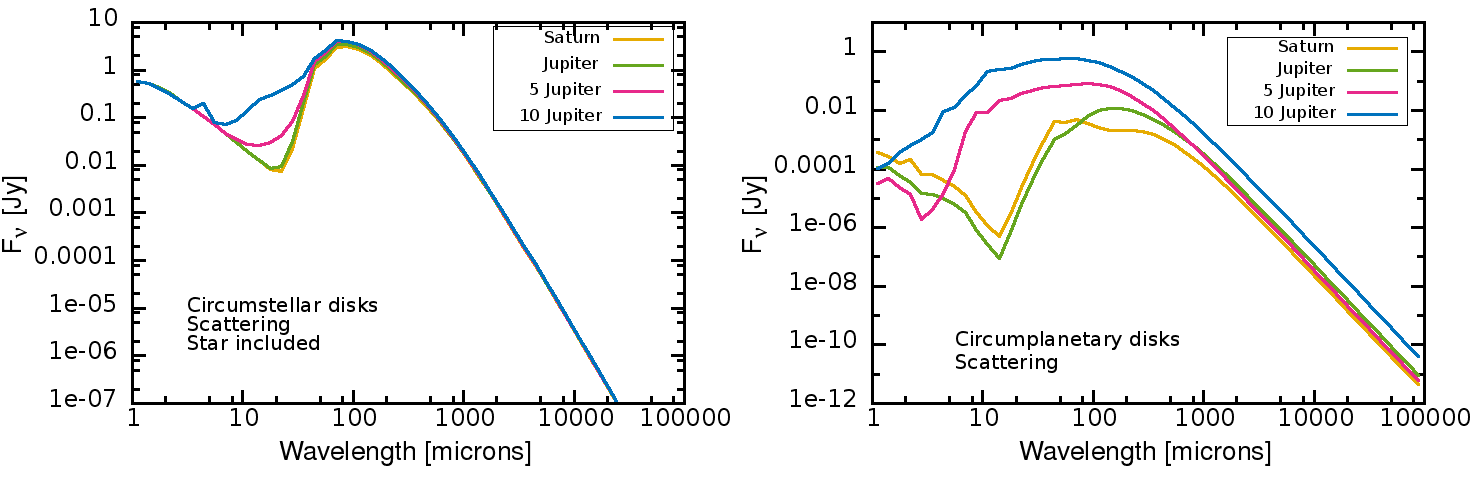}
\caption{Left: SED of the circumstellar disks (the disk ranges between 20-120 AUs). Right: SED of the CPDs. Both SEDS are scaled to 100pc distance.}
\label{fig:seds}
\end{figure*}

The hydrodynamic simulations contain an entire ring of the circumstellar disk between 20 and 120 AU, therefore it is possible to examine the spectral energy distributions of this disk. The resulting SEDs are shown in the left panel of Fig. \ref{fig:seds}. Because the CSD SEDs only have the contribution between 20 and 120 AU, they are like transitional disks with a large cavity.

On the right-hand panel of Fig. \ref{fig:seds} we show the CPD SEDs. Between the different planetary mass simulations, the largest difference of the CPD SEDs is in the 10-100 micron regime, but shorter and longer wavelengths also give some differences. In the near-IR ($<5 \mu m$), the difference in flux between the different planetary masses is minimal, and the flux here does not scale with the planetary mass. 

\begin{figure}
\includegraphics[width=\columnwidth]{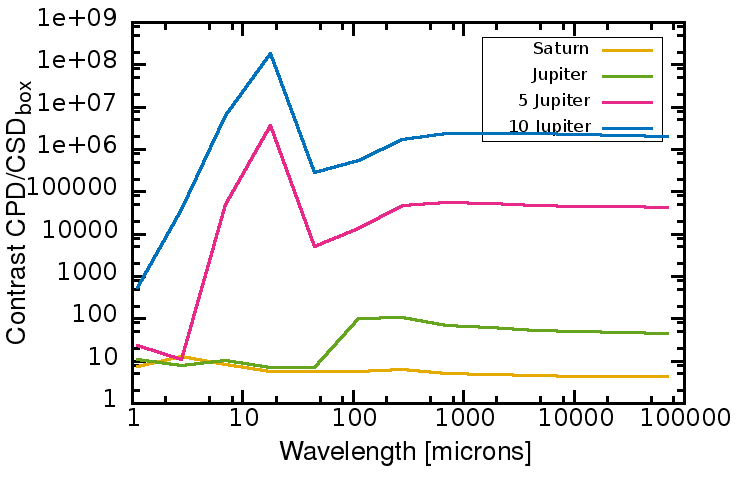}
\caption{The ratio between the flux of the CPD and the corresponding CSD flux defined in the same sized box (10\,AU $\times$ 10\,AU) at the "anti-planet" location (180 degrees azimuthally from the planet position) in function of the wavelength. Both SEDs have flux contribution within these boxes laying between 45 and 55 AU from the star, because the planets are at 50 AU.}
\label{fig:sed_contrast}
\end{figure}

Our science goal is to detect the CPD on the background of the CSD, therefore the contrast of their SEDs can inform us which wavelength-range is the best to detect the CPD. Since the CPD SEDs were evaluated only in a small area around the planet (in a 10\,AU $\times$ 10\,AU box, radially between 45 AU and 55 AU from the star, where the planet is at 50 AU), we determined the CSD SED within the same sized area as well, 180 degrees away from the planet azimuthally. We divided the two box SEDs with each other, to see the contrast between the CPD and CSD SEDs. The resulting contrast informed us about the wavelength-range where the difference between the CPD and the CSD is the largest (Fig. \ref{fig:sed_contrast}). For all models, the contrast is poor in the near-IR. For the gap-opening, 5 and 10 $\rm{M_{Jup}}$ planets the contrast is the best between $\sim$ 8 and 33 microns ($10^{7}$ and $10^9$, respectively), so in the vicinity of the 10-micron silicate feature. This is because the CPD is denser than the circumstellar disk in the large vicinity of the planet, and in addition, the CSD-box is centered at the anti-planet location, i.e. mostly capturing the planetary gap. It is of course easier to spot the CPD in a deep gap, than if the planet would not have opened a gap. The second favorable wavelength-range for the highest contrast is beyond 400 microns up until 10cm, i.e. till the edge of the examined wavelength-scale. For the small mass planets, which could not open deep gaps, Fig. \ref{fig:sed_contrast} shows that a local maximum of the contrast is around 160-200 microns. Beyond this peak, the contrast is roughly constant till 10cm. The significance of the CPD scales with the planetary mass, the larger is the planetary mass, the easier is to detect the CPD in contrast to the CSD. 

Because the SEDs are impacted by the effects of dust-scattering and the inclusion of the star, we ran a number of tests to disentangle the different effects on the fluxes at the various wavelengths. We have made SEDs with and without scattering, with and without starlight, and ten-times enhanced dust-to-gas ratio (i.e. $10\times$ enhanced dust densities) to understand the impacts on the near-IR fluxes, but also to give prediction on all wavelengths between 1 microns to 10 centimeters. In the following sections these tests and their results will be described. 

\subsubsection{Effects of Scattering}
\label{sec:sed_test_scat}

Scattering on dust is an important effect below 10-20 microns. The SEDs show some features (e.g. spikes) at these near-, mid-IR wavelengths, however these spikes are just the result of the Monte-Carlo scattering within the RADMC-3D, not molecular/atomic lines. To produce SEDs, one has to limit the amount of photons used in the radiative transfer (RT) runs in order to have reasonable computational time. When running with the same amount of photons, but different seed number for the Monte-Carlo, the flux can change a little. Hence the spikes should not be taken seriously, or mistaken with gas lines (which were not included in these RT runs). Moreover, we have tested that the reasons for the photon noise lies in the fact that the hot gas-dust is very optically thick ($>100$) in the intermediate vicinity of the planet, hence only few photons can escape even if the calculation is done with $10^7$ photons. Simply this fact causes a lot of photon noise, that is inevitable.

\begin{figure*}
\includegraphics[width=18cm]{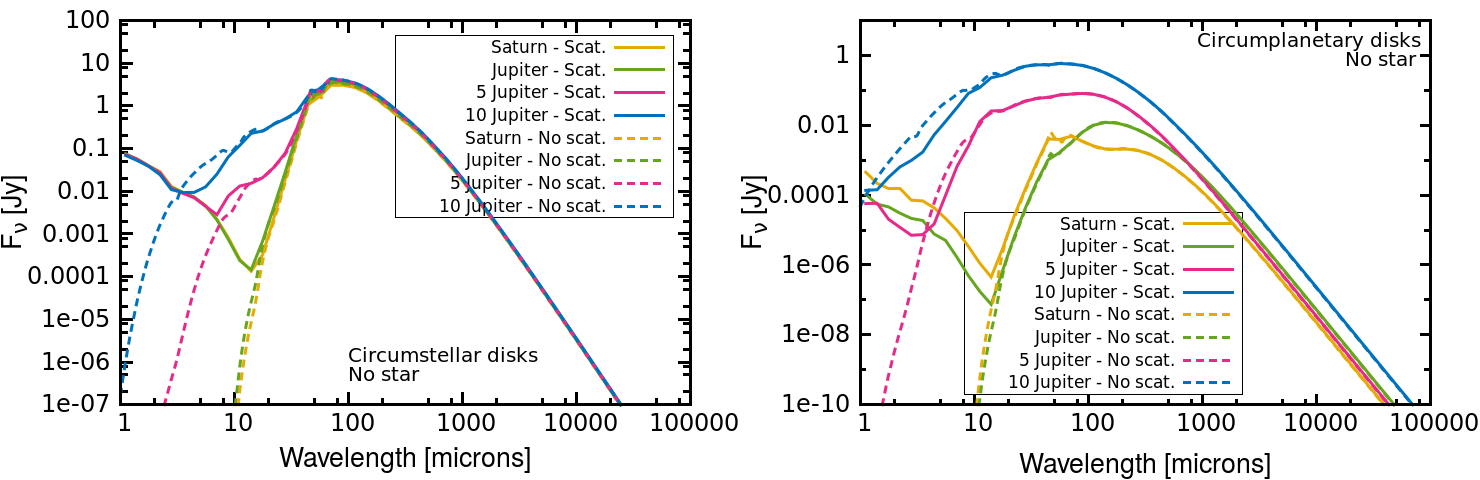}
\caption{Left: SED of the circumstellar disks with and without scattering; the star is not included. The CSD ranges between 20 AU till 120 in these simulations. Right: SED of the CPDs, comparison with and without scattering. Fluxes determined at 100 pc distance.}
\label{fig:seds_scat_noscat}
\end{figure*}

Nevertheless, the inclusion of the scattering gives significantly higher fluxes both for the CSD and the CPD (Fig. \ref{fig:seds_scat_noscat}). When not including any star in the model, the scattering gives basically all the flux for the CSD SED below 16 microns for Saturn and 1-Jupiter models (left panel in Fig. \ref{fig:seds_scat_noscat}). The turnover is a bit different at the larger mass planets, it gives most of the flux below 8 microns for the $5\,\rm{M_{Jup}}$ planet, and below 3 microns for the $10\,\rm{M_{Jup}}$ case (left panel in Fig. \ref{fig:seds_scat_noscat}). In the case of the CPD SED, scattering contributes basically all the flux below 10 microns, except for the $10\,\rm{M_{Jup}}$ model, where the no scattering run gives a bit higher flux (right panel in Fig. \ref{fig:seds_scat_noscat}). Beyond 10-20 microns the fluxes are basically the same with or without scattering. 

\begin{figure}
\includegraphics[width=\columnwidth]{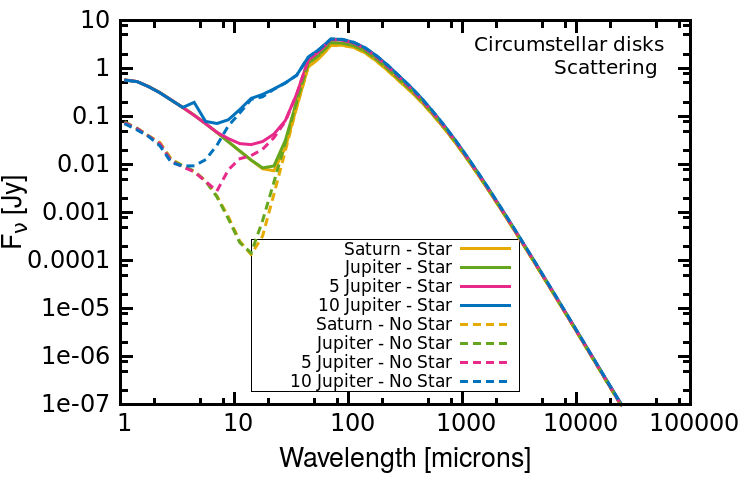}
\caption{Circumstellar disk (ranges between 20 and 120 AU) SED with scattering, with and without the inclusion of the star. Scaled to 100 pc distance.}
\label{fig:seds_scat_star_nostar}
\end{figure}

If we include scattering and compare the same models with and without inclusion of the star, we get of course some difference for the SED of the circumstellar disk (Fig \ref{fig:seds_scat_star_nostar}). It has larger flux below 25 microns if the star is included (stellar spectrum's Rayleigh-Jeans tail) for Saturn- and Jupiter-models, the other two cases have transition a bit earlier, around 10 microns. Beyond 80 microns all the SEDs are basically the same, since at this part the stellar irradiation dominates. The inclusion or absence of the star of course does not affect the CPD SED, because the CPD lies far away from the star, the box where the flux was defined does not include the star (and the heating from the planet remains the same and this effect dominates in the CPD).

\subsubsection{Effects of Enhanced Dust-to-Gas Ratio}
\label{sec:sed_test_dust}

To understand how the flux changes with dust disk mass, we enlarged the dust density in every cell by a factor of 10. Hence both the CSD and the CPD increased its dust mass by a factor of 10. This way the dust-to-gas ratio became 10\% instead of 1\%.

\begin{figure*}
\includegraphics[width=18cm]{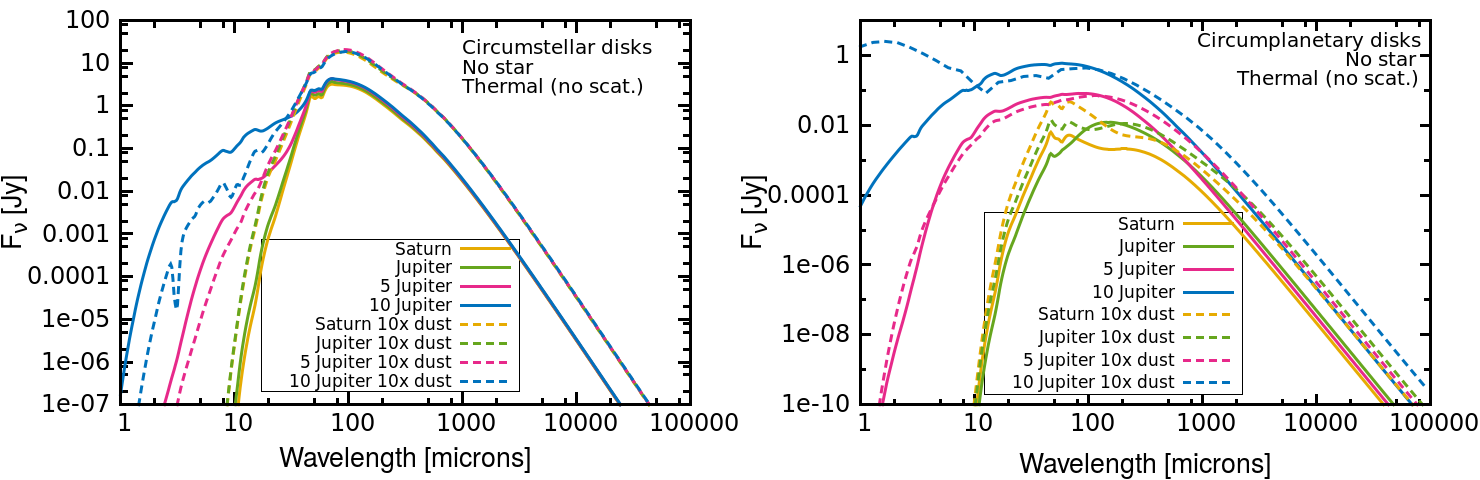}
\caption{Left: SED of the circumstellar disks (ranges between 20 and 120 AU) with and without 10x enhanced dust mass (dust-to-gas ratio of 10\%). There is no star, nor scattering included in this SED. Right: corresponding SED of the CPDs. Distance is  100pc.}
\label{fig:seds_dust_nodust_nostar_noscat}
\end{figure*}

First, we compare the SEDs of the nominal amount of dust with the enhanced dust, without star and without scattering (i.e. thermal radiation only). For the CSD, it reduces the flux below 10 microns for $5\,\rm{M_{Jup}}$ and $10\,\rm{M_{Jup}}$ planet models, for the smaller mass planets it enhances a bit (left panel in Fig. \ref{fig:seds_dust_nodust_nostar_noscat}). This is due to the difference in gap-opening, the two smaller mass planets do not open deep gaps. Beyond 20 microns it adds flux for all the models (at these wavelengths the CSD is optically thin). The CPD SEDs show similar features (right panel in Fig. \ref{fig:seds_dust_nodust_nostar_noscat}): in the $5\,\rm{M_{Jup}}$ and $10\,\rm{M_{Jup}}$ models there is a bit more flux in the near-, mid-IR, while the models with the smaller mass planets remain the same. 

\begin{figure*}
\includegraphics[width=18cm]{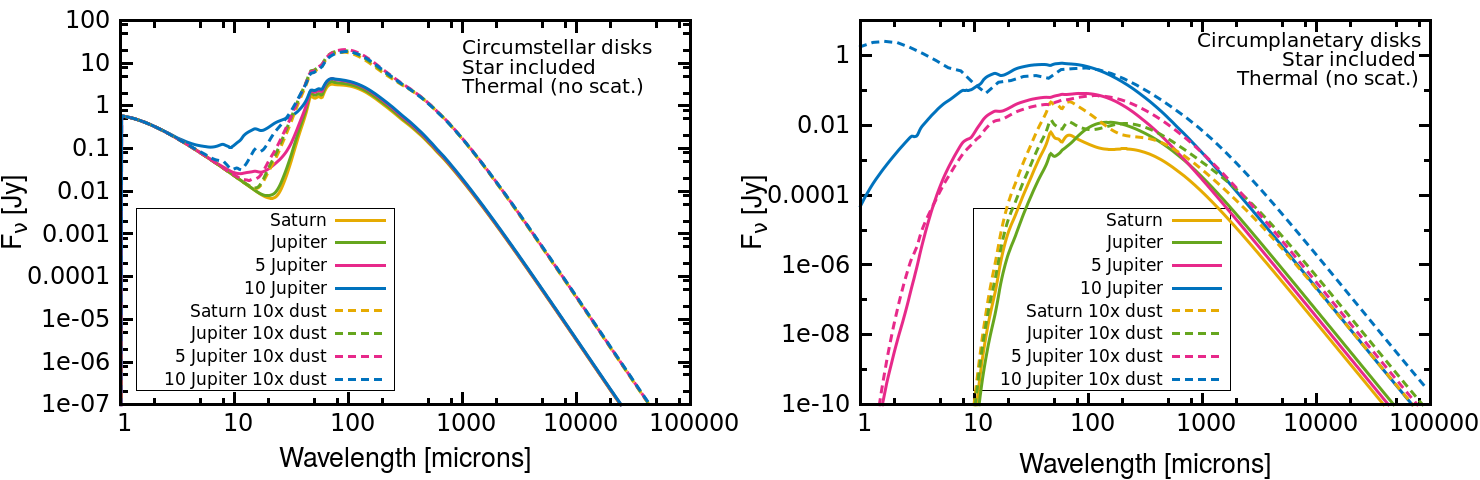}
\caption{Left: SED of the circumstellar disks (ranges between 20 and 120 AU) with and without 10x enhanced dust mass (dust-to-gas ratio of 10\%). The star is included, but scattering is not. Right: corresponding SED of the CPDs. Scaled to 100pc distance.}
\label{fig:seds_dust_nodust_star_noscat}
\end{figure*}

When repeating the same exercise (no scattering, with and without enhanced dust density) but adding the star, the result is very similar. The CSD SED is characterized with enhanced flux beyond $\sim$ 20 microns when the dust density is $10\times$ the nominal one (left in Fig. \ref{fig:seds_dust_nodust_star_noscat}). Between 5 and 20 microns it usually results in less flux when adding more dust to the models. Below $\sim$ 5 microns the flux is the same. The CPD SED contains a little more flux in the near-, mid-IR for the two large mass planet cases (right in Fig. \ref{fig:seds_dust_nodust_star_noscat}), the smaller mass planet models are roughly the same irrespective of the enhanced dust-to-gas ratio in the CPDs. 

\begin{figure*}
\includegraphics[width=18cm]{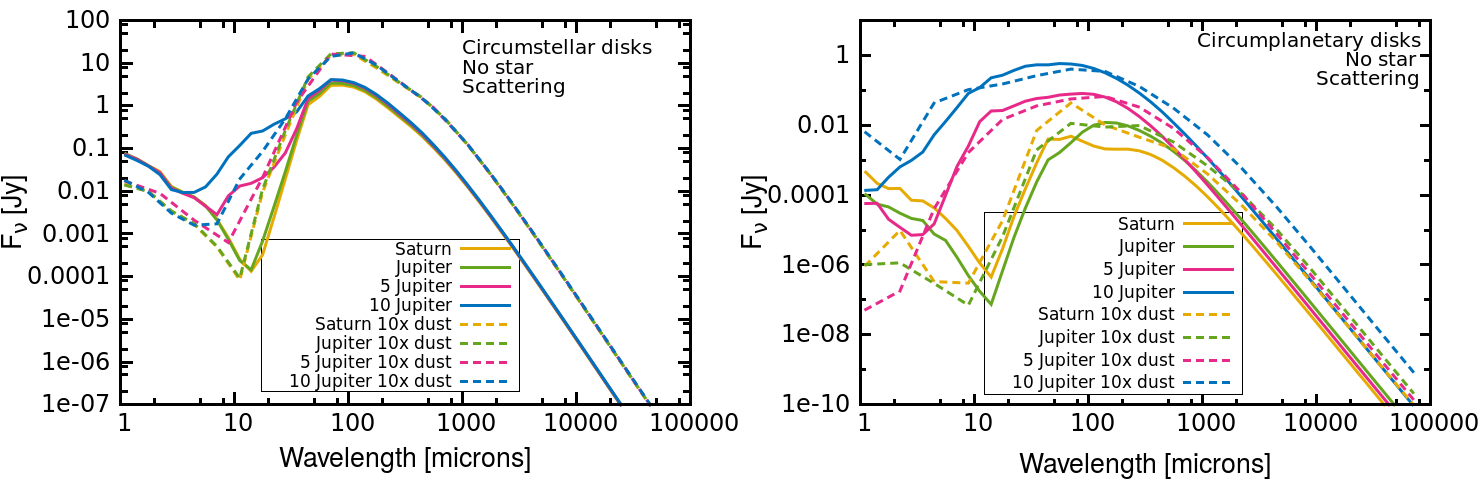}
\caption{Left: SED of the circumstellar disks (ranges between 20 and 120 AU) with and without 10x enhanced dust mass (dust-to-gas ratio of 10\%). The star is included, as well as scattering. Right: corresponding SED of the CPDs. Fluxes shown at 100pc distance for both SEDs.}
\label{fig:seds_dust_nodust_scat_nostar}
\end{figure*}

When one includes scattering for the 10x dust density models, and compares them with the nominal models with scattering the results are somewhat unexpected. In this case there is no star included for clarity (Fig. \ref{fig:seds_dust_nodust_scat_nostar}). The CSD SED has less flux below 25-40 microns (depending on the planet mass) if the dust density is 10x higher, however, there is always more flux beyond 50 microns with increased dust density (left panel in Fig. \ref{fig:seds_dust_nodust_scat_nostar}). The CPD SEDs are quite noisy in the near-IR due the photon noise, but for the Saturn-, Jupiter-mass and 5 Jupiter-mass planet models the inclusion of more dust reduces the flux below $\sim$ 10 microns, while one can expect the opposite for the $10 \rm{M_{Jup}}$ planet in this wavelength region (right panel in Fig. \ref{fig:seds_dust_nodust_scat_nostar}). Beyond 200 microns the 10x dust gives higher fluxes for all models. In all wavelengths the difference between the nominal and 10x dust mass models are very small. 

\subsection{Near-IR Fluxes}
\label{sec:nearIR}

\begin{figure}
\includegraphics[width=\columnwidth]{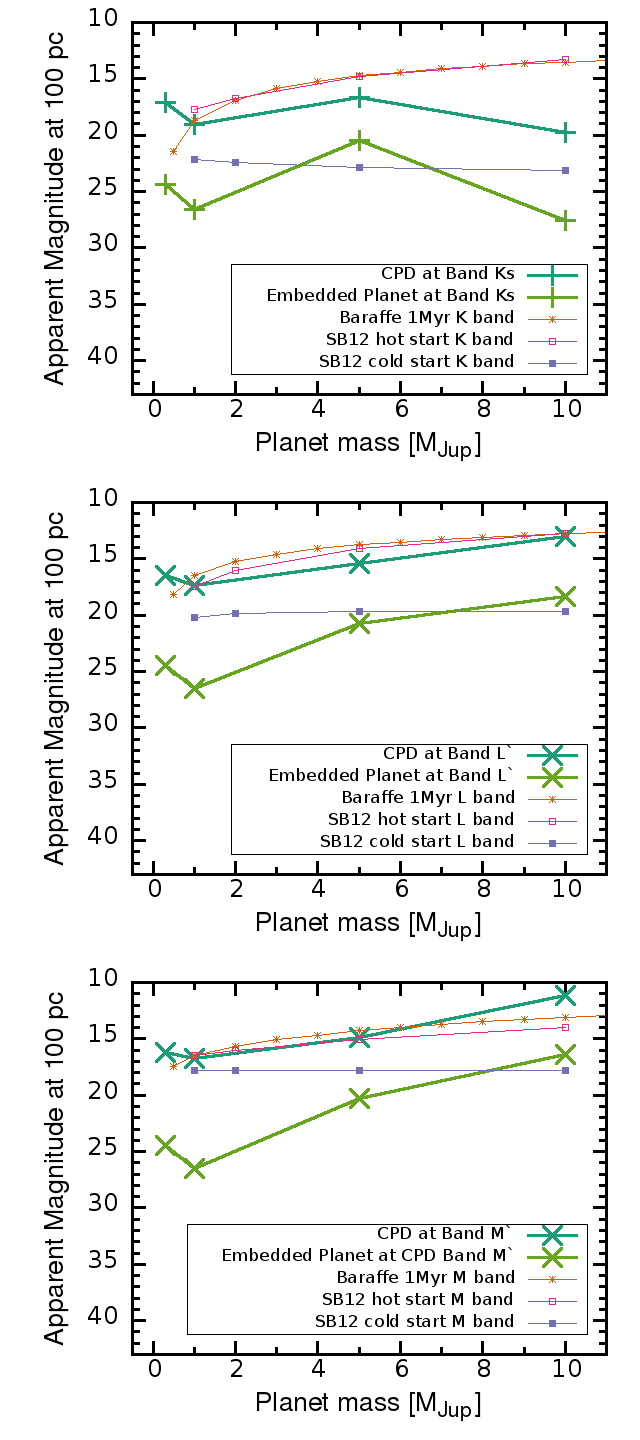}
\caption{Apparent magnitudes of Ks, L', M' bands of the embedded planet and the CPD+planet.}
\label{fig:magnis}
\end{figure}

In Table \ref{tab:fluxes} we summarize the fluxes and magnitudes of the embedded planet and the CPD+planet from our simulations in Ks, L', M' bands. The circumstellar disk contribution to the CPD fluxes was already subtracted. The Figure \ref{fig:magnis} shows visually the fluxes for each given band, and in comparison to theoretical models of \citet{Baraffe03} and \citet{SB13}. The CPD as a whole is brighter than the planet itself, which is also a consequence on how we determined their fluxes: for the CPD we integrated the flux of the pixels within one aperture and for the planet we only refer as a peak value within this region. Nevertheless, from the CPD aperture flux we subtracted the circumstellar contribution by defining the flux at the anti-planet location (i.e. 180 degrees away from the planet azimuthally) within one aperture as well. The ratio between the CPD flux (without the subtraction of the CSD flux) and the flux at the anti-planet location within one aperture is the $\rm{SNR_{CPD}}$ listed in the Table \ref{tab:fluxes}. This helps determining how significant is the CPD signal in comparison to the underlying CSD. This contrast is at least two orders of magnitude larger for the 5 and 10 $\rm{M_{Jup}}$ planet models, than in the lower planetary masses (last column in Table \ref{tab:fluxes}). In fact, when one visually inspects the synthetic images on Fig. \ref{fig:mocks}, the presence of the planet is not obvious for the Saturn and Jupiter cases, in these cases the detection of the hot spot due to the planet+CPD would be very unlikely.  

\begin{table}
\begin{centering}
  \caption{Embedded Planet and Circumplanetary Disk Fluxes and Magnitudes at 100 pc}
 \label{tab:fluxes}
  \begin{tabular}{llccccc}
  \hline
$\rm{M_{P}}$  & Band & $\rm{Flux_{P}}$ & $\rm{Flux_{CPD}}$& $\rm{Mag_{P}}$ & $\rm{Mag_{CPD}}$  & $\rm{SNR_{CPD}}\dagger$  \\
  \hline
 $[\rm{M_{Jup}}]$ &  &  [Jy] & [Jy] & [mag] & [mag] & \\
 \hline
10.0 & Ks & 4.22e-16 & 8.14e-06 & 27.56 & 19.76  &   227.81 \\ 
10.0 & L' & 7.84e-13 & 1.55e-03 & 18.35 & 13.03  & 10149.32 \\ 
10.0 & M' & 2.73e-12 & 5.38e-03 & 16.43 & 11.11  &  7528.27 \\ 
 5.0 & Ks & 3.00e-13 & 1.48e-04 & 20.43 & 16.61  &  1079.38 \\ 
 5.0 & L' & 8.70e-14 & 1.78e-04 & 20.74 & 15.38  &   141.01 \\ 
 5.0 & M' & 8.11e-14 & 1.70e-04 & 20.25 & 14.87  &    48.61 \\ 
 1.0 & Ks & 1.03e-15 & 1.57e-05 & 26.58 & 19.05  &     3.65 \\ 
 1.0 & L' & 4.38e-16 & 2.87e-05 & 26.49 & 17.36  &     3.78 \\ 
 1.0 & M' & 2.61e-16 & 3.00e-05 & 26.48 & 16.75  &     2.79 \\ 
 0.3 & Ks & 7.71e-15 & 9.10e-05 & 24.40 & 17.14  &    13.85 \\ 
 0.3 & L' & 2.72e-15 & 6.69e-05 & 24.50 & 16.44  &     4.92 \\ 
 0.3 & M' & 1.64e-15 & 4.85e-05 & 24.49 & 16.23  &     3.01 \\ 
\hline 
\end{tabular}
\raggedright{\tiny{$\dagger\rm{SNR_{CPD}}$ is defined as the flux ratio between the CPD flux within the aperture versus the flux of the CSD within the same sized aperture 180 degrees away from the planet, i.e. at the "anti-planet" location.}}
\end{centering}
 \end{table}

In comparison to the theoretical models of Baraffe and Spiegel \& Burrows, it is clear that the embedded planets in our simulations are fainter, because the extinction due to the CPD is significant. It is just a coincidence that some of the CPD magnitudes fit well to the planet brightness predictions of these evolutionary models. We did not try to match the luminosity of the planets in our simulations with either of the above planet evolution model prediction, because the planet luminosity in the formation phase is unknown. The evolutionary models either way do not include the hot-bath of the CPD in this evolutionary phase, their planets are always considered to be detached from their parent disk. To access the CPD/planet brightness ratio, therefore, we compare the planet and the CPD both from our simulations for consistency.  

\begin{figure*}
\includegraphics[width=18cm]{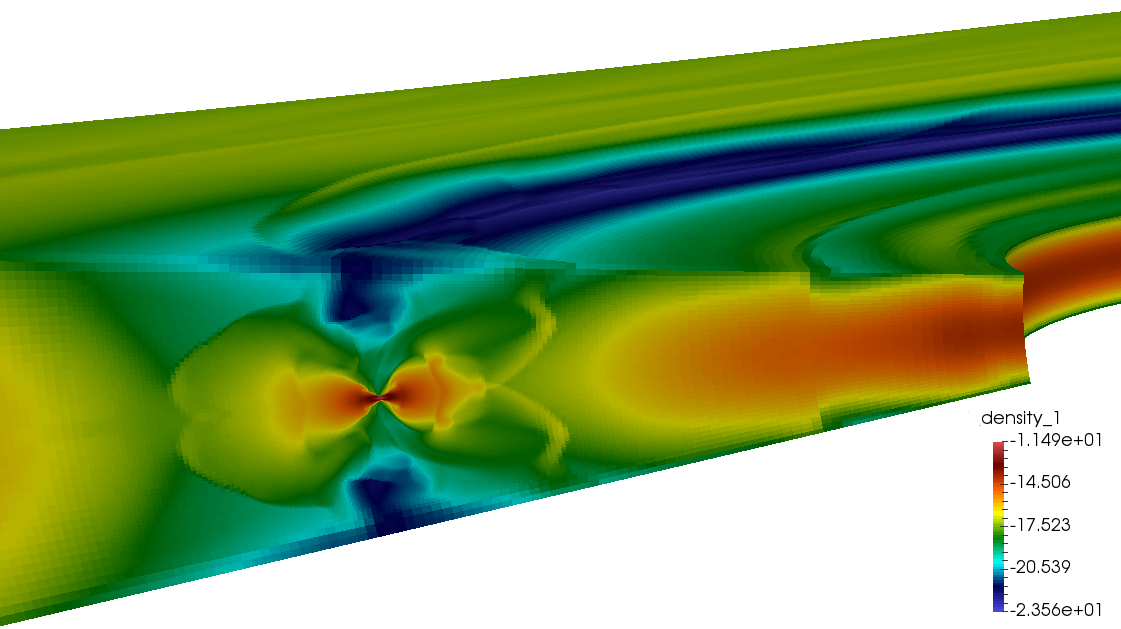}
\caption{The gas density from one of the hydrodynamic simulations in order to show how much the planet is embededded. The small disk with the orange color is the CPD, and the planet can be seen in the middle of that disk. The green colors show that there is significant amount of dust and gas density above the planet, even though the planet opened a gap, hence the extinction of the planet brightness is strong.}
\label{fig:extinction}
\end{figure*}

In Figure \ref{fig:magnis} one can see that the embedded planet- and CPD-magnitudes do not scale with the planetary mass. This is due to the fact that the scattering is important in this wavelength range. If one would consider purely the thermal (i.e. without scattering) fluxes, both the CPD and the planet fluxes would be ordered according to the planet masses, i.e. the higher mass planet has brighter CPD, as can be seen in the SEDs in the right panel of Fig. \ref{fig:seds_dust_nodust_scat_nostar}. As it was described in Sect. \ref{sec:sed_test_scat}, the Monte-Carlo scattering within the RADMC-3D runs can result in different fluxes on the CPDs, depending on what was the seed number, or how many photons were used. Hence, to get robust results on the Ks, L', M' magnitudes, we run the same RADMC-3D runs ten times with $10^7$ photons in each case, then took the median value in each pixel of the ten images, and calculated the brightness values in this final image. Hence the reported brightnesses are robust, the fact that the near-IR fluxes does not scale with planetary mass is not the effect of the Monte-Carlo noise.

The observability of the planet+CPD is strongly dependent on the planetary gap depths in the near-IR as well. Figure \ref{fig:mocks} shows that the gaps are clearer and deeper towards the longer wavelengths, making it easier to separate the CPD from the circumstellar disk. Of course, the more massive is the planet, the deeper and wider gap it creates within the circumstellar disk, hence observing larger mass planets is always easier. 

We have run several tests to understand the robustness of the fluxes. First, we tested the number of photons used in the RADMC-3D runs, whether the fluxes we determined above are converged. In case one uses too low amount of photons, the fluxes can change significantly. Therefore, instead of $10^7$ photons used for scattering in the nominal models, we also run a suite of RT modeling with an order of less photons, i.e. only $10^6$ photons for the Saturn-mass planet model. The error on the intensities were within 0.06\%, 0.01\%, and 0.04\% for Ks, L', M' bands, respectively. Moreover the the brightness of the CPD will depend on the dust albedo and on the optical thickness of the CPD. We describe the findings on the planet and CPD magnitudes in Sect. \ref{sec:plmass} in order to address the issue of the planetary mass estimation from the observed brightnesses.


\subsection{Planetary Mass Overestimation}
\label{sec:plmass}

\begin{figure}
\includegraphics[width=\columnwidth]{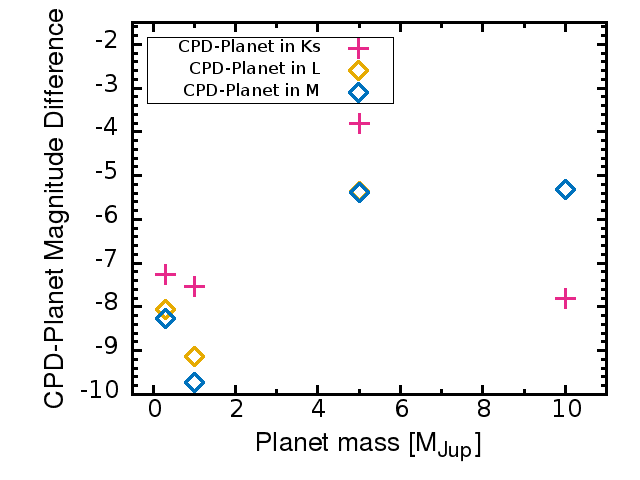}
\caption{The magnitude difference between the CPD and the embedded planet in our models at the Ks, L' and M' bands.}
\label{fig:mag_diff}
\end{figure}

In Sect. \ref{sec:nearIR} it was mentioned that the CPD flux is significantly larger than that of the planet. It is worthwhile to quantify this difference at the various bands, in order to give an estimate about how much the planetary mass from the observed flux can be over-estimated, by assuming that the detected flux solely comes from the planet, and not from the CPD. In Fig. \ref{fig:mag_diff} we show the difference between the CPD magnitude and the embedded planet from our models. For the barely gap-opening planets (i.e. Saturn and Jupiter) all three bands are giving roughly the same difference of $\sim$ 7.8 and $\sim$ 8.6 magnitude, respectively. This large difference can lead to a factor of 100 over-estimation of the planet mass according to the Baraffe-models (assuming an age of 1 Myr). For the larger mass planet cases ($5\,\rm{M_{Jup}}$ and $10\,\rm{M_{Jup}}$) in L' and M' bands the difference is 5.3 magnitudes, which translates to a factor of 20 over-estimation of mass using the Baraffe-models.

The brightness of the CPD will depend on the dust albedo and on the optical thickness of the CPD and hence its mass. In near-IR most of the flux is coming from the photon-scattering on dust particles, hence the amount of dust in the CPD will matter for the magnitudes. Because the CPD is continuously fed by the circumstellar disk, its mass mostly changes together with that of the CSD \citep{Szulagyi17gap}. Therefore we tested the fluxes when the dust density was increased by a factor of 10 in every cell of the models (i.e. both the CSD and the CPD mass was increased by an order of magnitude). We found that the CPD magnitude does not linearly scale with dust mass. The CPD brightness at 100 pc will be 3.8 magnitude fainter for the Saturn and Jupiter-mass planet models if the dust density is increased by 10. On the contrary, for the higher mass planets the brightness will increase, for the 5 Jupiter-mass case by 7.6 magnitude, for the $10\,\rm{M_{Jup}}$ case by 1.5 magnitude on average for the bands Ks, L', M'. This shows that the CPD brightness in near-IR will strongly depend on the dust mass, and whether or not there is a deep gas gap around the CPD (i.e. the amount of extinction). 


Our tests have revealed, that no matter what model we use, the CPD brightness will be larger than that of the planet's by several magnitudes. The CPD brightness will depend on the disk parameters, e.g. its the dust-to-gas ratio. Hence, when trying to estimate the forming planet mass based on near-IR observations, the planet evolutionary models cannot be used. These models can be useful when there is no more gaseous CPD around the planet, i.e. when there is no more gaseous circumstellar disk in the system. 

\section{Discussion}

Like every model and simulation, ours also have several limitations. First of all, in this work we used the assumption of a fixed dust-to-gas ratio of 0.01. It is known since a long while that CSDs can have dust traps where locally the dust-to-gas ratio is enhanced \citep[e.g.][]{YG05,Marel13,DD14,Birnstiel12,Meheut12}, but recently such traps has been found even in CPDs \citep{DSz18}. Even considering the bulk dust-to-gas ratio of the disks globally, different systems show very different values between 0.1-0.001 for CSDs \citep[e.g.][]{WB14,Ansdell16}. In Sect. \ref{sec:sed_test_dust} we ran tests about if the the dust-to-gas ratio is larger than 1\%. We expect, that if the ratio would be smaller than 1\%, then we would find the opposite behavior as described in Sect. \ref{sec:sed_test_dust}. Beyond 20-50 microns there would be less flux in the SEDs than our nominal 1\% case, and more in flux in the shorter wavelengths.

The semi-major axes of the planets will also play a major role on the CPD brigthness. The further away the planet lies from the star, its CPD will be colder just as a tumbs of rule. As the gas density is smaller in the outer CSD, the CPD will get optically thinner, hence its cooling time will enhance rapidly. On the other hand, if the CPD is optically thick (e.g. lies close to the star in the optically thick part of thr CSD, or when the CSD is massive), then it will be significantly hotter, hence brighter. Even though we have 5 AU simulations at hand, with the same aperture sizes it would have been impossible to measure the CPD and planet magnitudes. 

Our hydrodynamic simulations are disk simulations, without proper treatment for the planet. Our approach was to fix the radius and mass of the gas giants, but in the formation phase these are highly unknown parameters. These values affect the planet brightness the most, hence with different setup our embedded planet fluxes could change as well. If the mass of the CSD is larger, the accretion rate to the planet will also increase \citep{Szulagyi16b}. The enhanced accretion rate in turn will increase the CPD brightness through the accretional luminosity \citep{Zhu15,Szulagyi16a}. In our previous work of \citet{SzM16} we estimated the accretion luminosity of planets with orbital separation of 5.2 AU. They all turned out to be on the order of a few times $10^{-5} \rm{L_{Sun}}$. We also calculated the accretional luminosities for the 50 AU separation planets presented in this work. The accretion luminosities are a few times $10^{-6} \rm{L_{Sun}}$, so smaller than the close-in planet's. In comparison, the Baraffe-model's planet luminosities at age of 1 Myr for a Jupiter-mass giant is almost $10^{-5} \rm{L_{Sun}}$ and for a 10 Jupiter-mass planet is $10^{-3} \rm{L_{Sun}}$. If these estimations are correct, the accretion luminosities of the planets (not the CPD accretion rate and the corresponding luminosity) are slightly smaller than the intrinsic planet luminosities. Note, that the planet accretion rate, and the corresponding accretional luminosity is smaller than the CPD accretion rate and the corresponding accretion luminosity. The accretion to the CPD is $\sim$ an order of magnitude higher than to the planet, and henceforth the accretion luminosity ($GM_{planet}\dot{M}_{planet}/R_{planet}$) of the CPD is also higher than that of the planet's. Due to the meridional circulation and that the CPD is a de-accreting disk, most of the gas entering the CPD will not end up in the planet, but will be reprocessed into the circumstellar disk \citep{Szulagyi14,FC16}.

Our work was not including hydrogen dissociation and ionization. This can have an affect on the temperatures and on the calculated magnitudes. If the energy would go into dissociating and ionizing hydrogen, the temperatures could be lower. However, the area in the simulation where the temperature would go as high to dissociate or ionize hydrogen, is extremely small, only few cells in these simulations. In comparison, when the planet was placed at 5.2 AU from the star, and therefore the CPD was in an optically thick and hotter part of the CSD, the dissociation and ionization only affected the very vicinity of the planet, inside of the planet and the hottest part of the shock front on the CPD surface (Fig. 3 in \citealt{SzM16}). 

Moreover, we neglected magnetic fields. This can change the accretion flow and rate to the CPD and to the planet \citep{OM16,Gressel13,KW15}, that could change our conclusions. Self-gravity was also neglected due to the small mass of CPDs. However, in heavy circumstellar disks, maybe the CPDs could also become gravitationally unstable \citep{LM12}.

As we showed it in Sects. \ref{sec:sed_test_dust} and \ref{sec:plmass} the dust mass of the CPD and the semi-major axis of the planet will affect the brightness of the planet and its CPD in all wavelength hence to determine the CPD brightness one should perform system-specific modeling. In addition, in this work we only considered one initial CSD mass and viscosity, we could not explore the parameter space further with these complex simulations.

\section{Conclusions}

We performed a study on the circumplanetary disk (CPD) brightness by creating their spectral energy distributions (SEDs) and by examine their near-IR (Ks, M', L') fluxes with synthetic images. We ran radiative hydrodynamic simulations of Saturn, Jupiter, 5 Jupiter- and 10 Jupiter-mass planets placed at 50 AU from their star still embedded in their natal circumstellar disks. In this phase, these planets are surrounded by a circumplanetary disk, which we examined in high resolution placing nested meshes in the planet vicinity. We then post-processed these hydrodynamic simulations with RADMC-3D radiative transfer tool to create wavelength-specific images and SEDs. For the dust, we assumed a constant 1\% dust-to gas ratio with a composition of 40\% silicate, 40\% water and 20\% carbon and we took care of the evaporation of these species if the temperatures were high enough in a given cell. 

Our finding is that the best contrast ratio between the CPD with the surrounding circumstellar disk is between $\sim$ 8 and 33 microns if the planets opened deep gaps (i.e. $\ge 5 \rm{M_{Jup}}$), and this contrast is particularly poor in the near-IR. For all planetary masses, the sub-mm/radio wavelengths provide also good contrast with the circumstellar disk, so this wavelength range provide unprecedented opportunity to detect forming planets and their circumplanetary disks, similarly as previous studies pointed also out \citep{Szulagyi17alma,IsellaTurner16}. The CPD SEDs revealed that the CPD brightness below $\sim$ 10 microns is dominated by the scattering of photons on the dust of this disk. In conclusion, to separate the CPD from the CSD observationally, the best chance is targeting sub-mm/radio wavelengths or in case of massive planets ($\ge 5 \rm{M_{Jup}}$) the 10-micron silicate feature vicinity.

We also pointed out that the CPD brightness is always higher than that of the embedded planet, which could easily lead to a planet-mass-overestimation by an order of magnitude. Hence, when trying to estimate the forming planet mass from the detected brightness, the planet evolutionary models cannot be used. In the formation phase we cannot detect the planets directly, only their circumplanetary disk. But the CPD magnitude will depend mainly on the disk parameters, e.g. the dust-to-gas ratio of the CPD, the distance from the star (CPDs closer to the star are hotter \& brighter), viscosity etc. We found that the CPD brightness does not scale linearly with the dust mass, and will strongly depend on whether the planet opened a deep gas gap ($\ge 5 \rm{M_{Jup}}$) or not ($\le 1 \rm{M_{Jup}}$) because of extinction. Our recommendation is to run system specific disk modeling if one is trying to estimate the planetary mass in the formation phase from the observed brightness.

\section*{Acknowledgments}

This work has been in part carried out within the Swiss National Science Foundation (SNSF) Ambizione grant PZ00P2\_174115. SPQ acknowledges the funding from the National Centre for Competence in Research ``PlanetS"  supported by  the  Swiss  National Science Foundation. Computations have been done on the ``M\"onch" machine hosted at the Swiss National Computational Centre.

\label{lastpage}

\end{document}